# Large diameter millimeter-wave low-pass filter made of alumina with laser ablated anti-reflection coating

RYOTA TAKAKU,[1] QI WEN,[2] SCOTT CRAY,[2] MARK DEVLIN,[3] SIMON DICKER,[3] SHAUL HANANY,[2,*] TAKASHI HASEBE,[4] TERUHITO IIDA,[5] NOBUHIKO KATAYAMA,[4] KUNIAKI KONISHI,[6] MAKOTO KUWATA-GONOKAMI,[1,6] TOMOTAKE MATSUMURA,[4] NORIKATSU MIO,[1,6] HARUYUKI SAKURAI,[7] YUKI SAKURAI,[4] RYOHEI YAMADA,[1] AND JUNJI YUMOTO[6]

[1] *Department of Physics, The University of Tokyo, 7-3-1 Hongo, Bunkyo-ku, Tokyo 113-0033, Japan*
[2] *School of Physics and Astronomy, University of Minnesota, 115 Union St. SE, Minneapolis, MN 55455, USA*
[3] *University of Pennsylvania, 209 S. 33rd St, Philadelphia, PA 19104, USA*
[4] *Kavli Institute for the Physics and Mathematics of the Universe (WPI), The University of Tokyo, 5-1-5 Kashiwa-no-Ha, Kashiwa, Chiba 277-8583, Japan*
[5] *ispace, inc., 3-42-3 Nihonbashi-Hamacho, Chuo-ku, Tokyo, 103-0007, Japan*
[6] *Institute for Photon Science and Technology, The University of Tokyo, 7-3-1 Hongo, Bunkyo-ku, Tokyo 113-0033, Japan*
[7] *Institute for Solid State Physics, The University of Tokyo, 5-1-5 Kashiwa-no-Ha, Kashiwa, Chiba 277-8583, Japan*
[*] *hanany@umn.edu*

**Abstract:** We fabricated a 302 mm diameter low-pass filter made of alumina that has an anti-reflection coating (ARC) made with laser-ablated sub-wavelength structures (SWS). The filter has been integrated into and is operating with the MUSTANG2 instrument, which is coupled to the Green Bank Telescope. The average transmittance of the filter in the MUSTANG2 operating band between 75 and 105 GHz is 98%. Reflective loss due to the ARC is 1%. The difference in transmission between the s- and p-polarization states is less than 1%. To within 1% accuracy we observe no variance in these results when transmission is measured in six independent filter spatial locations. The alumina filter replaced a prior MUSTANG2 Teflon filter. Data taken with the filter heat sunk to its nominal 40 K stage show performance consistent with expectations: a reduction of about 50% in filters-induced optical power load on the 300 mK stage, and in in-band optical loading on the detectors. It has taken less than 4 days to laser-ablate the SWS on both sides of the alumina disk. This is the first report of an alumina filter with SWS ARC deployed with an operating instrument, and the first demonstration of a large area fabrication of SWS with laser ablation.

## 1. Introduction

Alumina's high thermal conductivity and transmission spectrum make it an appealing low-pass filter material for instruments operating at millimeter wavelengths [1,2]. It is absorptive above a frequency of ∼1 THz, but highly transmissive at lower frequencies [3,4]. Near a frequency of 100 GHz (wavelength of 3 mm) it has low absorption loss at both room and cryogenic temperatures [5], and it has thermal conductance higher by factors of hundreds relative to plastic based materials [6] that are also used for making low-pass filters. Alumina is already used as an absorptive low-pass filter [2], and is planned to be integrated into future instruments [7,8].

One challenge in using alumina as a filter material is its high index of refraction $n = 3.12$ [3]. The high index of refraction leads to high reflection loss. The average reflectance across a 30% fractional bandwidth of a 2 mm thick slab of alumina at 100 GHz is 45%. Reduction of reflection loss is achieved by applying an anti-reflection coating (ARC) to the light entry and exit surfaces of the filter. A number of publications discuss ARC approaches for alumina [9–15].

A useful ARC approach is to fabricate subwavelength structures (SWS) on the native alumina material [16,17], creating a layer with an intermediate effective index of refraction $n_{\text{eff}}$ between the ambient environment of the filter, which typically has $n = 1$, and the material. Different SWS topologies give rise to different refractive index profiles [18–20]. The approach is particularly effective when the filter needs to operate at cryogenic temperatures because there is no differential contraction between the ARC and the substrate material. Because alumina ranks 9 on the Mohs Hardness Scale, standard machining of the material is challenging; Nitta et al. [21] report SWS made with dicing blades on alumina samples of unknown size. To alleviate mechanical machining challenges we have developed an approach using laser ablation [11,22]. Using picosecond and femtosecond lasers we fabricated SWS ARC structures on alumina, sapphire, and silicon, and showed that fabricated structures match designs both in shape and transmission properties [22–28]. With sapphire we demonstrated 116% fractional bandwidth with 97% transmission between 43 and 161 GHz [23]. With alumina we recently reported an ablation rate of up to 34 $mm^3$/min for structure heights near 1 mm, the highest ablation rate yet reported for these types of structures [25]. Nitta et al. [9] demonstrated laser ablation of a 1 cm square sample of alumina with YAG laser with an ablation rate of 0.6 $mm^3$/min.

All previous alumina, sapphire, and silicon samples with laser ablation-based SWS had diameters of 10 cm or less and were fabricated as part of technology development. To our knowledge, no alumina optical element with SWS ARC has yet been implemented with an operating instrument. In this paper we report the first fabrication of a laser-ablated alumina-based filter to be used with a currently operating millimeter-wave instrument. The instrument is MUSTANG2, which is coupled to the 100 m Green Bank Telescope (GBT) [29]. The filter is 30 cm diameter and to our knowledge is the largest optical element that has been laser-ablated with SWS ARC to date.

We give a brief review of MUSTANG2 in Section 2, discuss the design of the filter and pre-fabrication validation in Section 3, describe the fabrication approach in Section 4, report on characterization of the filter including index, loss, shape and mm-wave transmission measurements in Section 5, and discuss and summarize the findings in Sections 6 and 7.

## 2. MUSTANG2

MUSTANG2 [29] is a bolometric-based imaging instrument that has been operating on the GBT since 2016. It has a single broad-band frequency with a band-pass of 75 to 105 GHz, angular resolution of 9”, field-of-view of 4.2′, and it can map a sky region of 6′ in diameter to 73 $\mu$K_RJ RMS in an hour. The instrument is being used for diverse science goals including observations of the Sunyaev-Zel’dovich Effect in galaxy clusters to obtain masses [30] and to search for substructure [31,32], surveying the Galactic plane [33], measuring the properties of dust in star-forming regions [34,35], and mapping emission from the Moon [36].

MUSTANG2 has an array of 217 transition edge sensor (TES) bolometers cooled to 300 mK with a helium-3 adsorption refrigerator. Space constraints on the GBT limit the diameter of the MUSTANG2 cryostat, which leaves little room for the reimaging optics typically used in millimeter-wave instrumentation. Instead, the array is placed at the secondary focus of the GBT. The bolometers are coupled to free space with profiled feedhorns. The array of horns has an optically active diameter of 250 mm and is 150 mm from the ambient temperature receiver vacuum window.

Filters between the vacuum window and the focal plane array block most of the 180 W of incident radiation. Cryogenic considerations limit the maximum optical load on the array to 5 $\mu$W. The filtering is achieved with a combination of reflective [37] and absorptive filters. Prior to the summer of 2021 MUSTANG2 had two absorptive filters, one made of polytetrafluoroethylene (Teflon) heat sunk to 40 K, and the other made of nylon heat sunk to 4 K. These plastics are cheap, readily available, and their low index permits use of a single layer of SWS ARC made by cutting grooves into their surfaces. However, their low thermal conductivity led to significant heating at their centers. Modelling and measurements showed that the center of the Teflon filter attained a temperature of 110 K, leading to significant re-radiated power, reducing filtering efficiency and increasing thermal load on the colder 4 K, 1 K, and 0.3 K stages. The predicted thermal load from the filters on the 0.3 K stage was 2.6 $\mu$W, about half of the total allowed, and the predicted in-band optical load on the detectors was 11 pW. Replacing the Teflon filter with alumina was predicted to lower the temperature of the 4 K nylon filter from 27 K to 12 K, reduce the filter thermal loading on the 300 mK stage to 0.7 $\mu$W, and lower the in-band detector optical loading due to the filters to 5 pW.

## 3. Design and validation

The filter is made from a disk of 99.5% pure alumina, model A995LD procured from NTK Ceratech. Its diameter is 302 mm, which exceeds the required optically active diameter of 284 mm. We require average in-band transmission of at least 98% at cryogenic temperatures, which implies up to 2% loss due to absorption and reflection.

The thickness of the filter is determined through a trade-off between increasing in-band transmission, ensuring out-of-band high frequency absorption, minimizing its in-band emission, and considering the filter's mechanical robustness. Mechanical robustness suggests that the filter thickness should be more than a few mm. Increasing in-band transmission and out-of band absorption demand thinner and thicker material, respectively. Reduction of in-band emission requires high thermal conductance, which increases with thickness. We discuss this trade-off in the next section.

### 3.1. Absorption and filter temperature

For the purposes of filter design we assume that, at the operating temperature of 40 K, the loss tangent would be $\tan\delta \leq 4 \times 10^{-4}$ [38]. With this loss tangent, the MUSTANG2 75-105 GHz band-average absorption of a 3, 4, and 5 mm thick alumina sheet is 0.7, 0.9 and 1.2%, respectively.

To assess the effect of thickness on filter temperature we use ThermalDesktop$^{\text{TM}}$ and find the temperature profile of the filter as a function of radius with a simplified physical layout of the MUSTANG2 configuration. We assumed the following three layer stack: the upper and lower layers are infinite black body sheets maintained at 300 K and 4 K, respectively; the intermediate layer is a 302 mm diameter black body disk with its edge heat sunk to 40 K. The disk has variable thickness and its temperature as a function of radius is allowed to come to equilibrium with the other two sheets. The thermal conductance as a function of temperature of the intermediate layer was programmed to be that of alumina [39]. We find that with 3 (5) mm thick alumina the central disk's temperature is 45.5 (43.0) K. The advantage of alumina as mm-wave filter material [1] manifests itself well when comparing this result to Teflon. The central temperature calculated for a 10 mm thick Teflon disk [40], which is 2-3 times thicker than the alumina disk, is 250 K.

The MUSTANG2 team carried out simulations that include all the reflective and absorptive filters along the optical path, the temperature stages to which the filters are heat sunk, and the emissivity and thermal conductance properties of Teflon or alumina. The simulations give a predicted central temperature of 110 K for 10 mm-thick Teflon, and 41 K for 3 mm-thick alumina when it replaces the Teflon.

Given the in-band absorption and filter temperature considerations, we have chosen an alumina disk thickness of 5 mm. Each layer of the SWS ARC is about 1 mm (see next section) leaving a solid mid-layer thickness of 3 mm, which would give an adequate thermal performance. In-band average absorption is expected to be 0.9% arising from an effective thickness of 4 mm. This effective thickness is due to the 3 mm mid-layer thickness and an additional 1 mm that is due to the two SWS layers; each is 1 mm tall with about 50% of material removed.

Inoue [41] measured the loss tangent of alumina at 30 and 300 K between 0.325 and 1.5 THz. The data for 99.5% pure alumina show approximately linearly increasing loss tangent with frequency above 0.6 GHz, with $\tan\delta = 1.5$ and $4.3 \times 10^{-3}$ at 1.5 THz and 30 and 300 K, respectively. Extrapolating the 30 K data to 5 THz gives transmission of less than 0.1% for 4 mm thick sheet. Fourier transform spectroscopy data at 300 K for 99.6% pure alumina [42] show strong absorption with $0.02 \leq \tan\delta \leq 1$ between 10 and 30 THz. Assuming a factor of three lower absorption at 30 K, as measured at lower frequencies [41], the transmission of 4 mm thick alumina sheet at 10 THz is less than 0.01%. In comparison, with 10 mm thick Teflon we estimate a transmission of more than 30% up to a frequency of 14 THz.

### 3.2. Reflection

The required in-band average transmission of 98% together with the expected absorption loss implies that average reflection loss across the band should be ≤ 1%. The design considers two key drivers: (i) setting the pitch of the SWS such that the onset of diffraction occurs at frequencies higher than the band, and (ii) setting the shape of the SWS to provide low-reflection and flat response over the 75 - 105 GHz band. The design of the SWS only serves as a guide for planning the fabrication and as an approximation for the final shape. Ultimately there is a required level of reflection across the band, but no requirement on producing a specific SWS shape.

To avoid higher-order diffraction, the largest pitch is given by [23,43]

$$p \leq \frac{c/\nu}{n + \sin\theta_i}, \tag{1}$$

where $c$ is the speed of light, $\nu$ is the frequency, $n$ is the refractive index of the material, and $\theta_i$ is the incidence angle. Taking $\nu = 115$ GHz, a value that is 10 GHz higher than the upper edge of the passband, $n = 3.13$ [41], and $\theta_{i,\max} = 14.5°$ from the receiver optics, we determined that $p \leq 770\ \mu$m.

The design of the shape is based on a two-layer approach motivated by obtaining low constant reflection across the band [44]. A single optimal quarter-wave thick layer can not satisfy the average reflection requirement. The indices of the two layers and their respective quarter-wave depths, all obtained using impedance matching theory, are given in Table 1. The resulting shape is shown in Fig. 1. It is not necessary and is inefficient to laser-ablate the material and obtain the vertical walls of the two layer design. To provide guidance for the transmission properties of the final structure we modify the vertical walls to become slanted and thus provide an effective index gradient. For each layer the gradient is inspired by a Klopfenstein profile [45] in the following way: a Klopfenstein index profile is generated between $n = 1$ and $n = 3.1$ with characteristic parameters $\Gamma_m$ and is offset by $n_{\text{off}}$; the parameters for each layer are given in Table 1 as well as the average indices $\bar{n}$, which are close to the indices for each layer in the initial two layer design.

**Table 1. Parameters for the two layer design.**

| | Index | Depth [$\mu$m] | Width [$\mu$m] | Pitch [$\mu$m] | Klopfenstein parameters |
|---|---|---|---|---|---|
| Layer I | 1.33 | 626 | 385 | 770 | $\Gamma_m = 0.5$, $n_{\text{off}} = -0.5$, $\bar{n} = 1.27$ |
| Layer II | 2.35 | 354 | 680 | | $\Gamma_m = 0.4$, $n_{\text{off}} = 0.5$, $\bar{n} = 2.28$ |

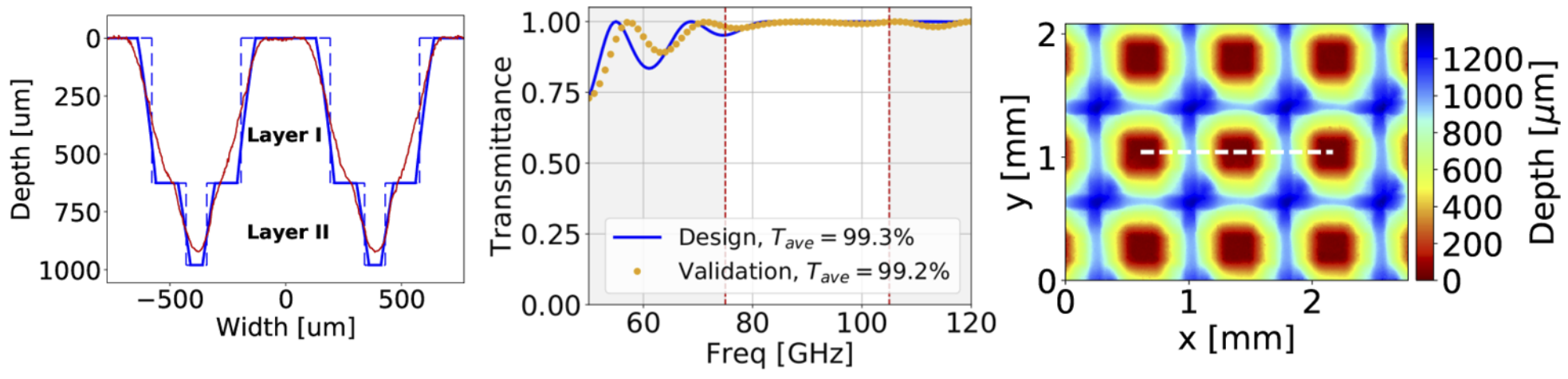

(a) (b) (c)

**Fig. 1.** (a) Cross sectional view of the SWS baseline design (blue) and of the validation sample (red). The design begins with a step structure consisting of two distinct layers (dash blue) and we modify it to produce the final baseline design that has slanted, curved walls (solid blue). Structure dimensions are given in Table 1. For the validation sample we show the cross-section marked in Panel c. (b) Predicted transmission of the final baseline design (blue solid), and of the validation sample (orange points). To illustrate the efficacy of the SWS ARC, the predicted transmission assumes no absorptive loss. (c) Top view of a section of the validation sample with a line showing the location of the profile cross-section in Panel a.

The conversion of the vertical walls to slanted is not unique and is done iteratively to ensure that the expected average reflection criterion is satisfied and that the slant angle of the walls is not more than 82 degrees from the horizontal, a slant we have already achieved in a past fabrication [11]. The index gradients are converted to physical shapes using 2nd-order effective medium theory [19] and the resulting shape is shown in Fig. 1. For ease of reference, we refer to the structure in each unit cell a ‘pyramid’.

Panel (b) of Fig. 1 shows a transfer matrix method (TMM)-based [46,47] expected transmission of the baseline design excluding absorption. The TMM calculation uses 1000 layers along the gradient of the index profile. The average reflection loss is 0.7%. We emphasize that we did not expect the fabricated structures to attain the two-stepped pyramid-like structure of the baseline design. Rather, the baseline design was used as a guide for planning the fabrication process, and we intended to assess the resulting shapes based on their transmission performance and iterate if necessary. It turned out that no iterations were necessary; see the next Section.

### *3.3. Validation sample*

Figures 1(c) and 2 show a section from a 5 × 5 pyramid sample we fabricated to serve as validation of the baseline design. Table 2 gives averages of geometric parameters of nine pyramids, and a cross-section along the line marked in Fig. 1(c) is shown in Fig. 1(a). Figure 1(b) shows the expected transmission of the MUSTANG2 filter if it were fabricated with the shape of the validation sample. We show the average of the two polarizations. The transmission is calculated using rigorous coupled wave analysis (RCWA) into which we programmed the averaged measured shape. The calculation ignores absorption. The average in-band transmission is 99.2%, implying 0.8% reflection, and the transmission spectrum has similar features to that expected for the baseline design. We therefore decided to maintain the same fabrication parameters for the full filter.

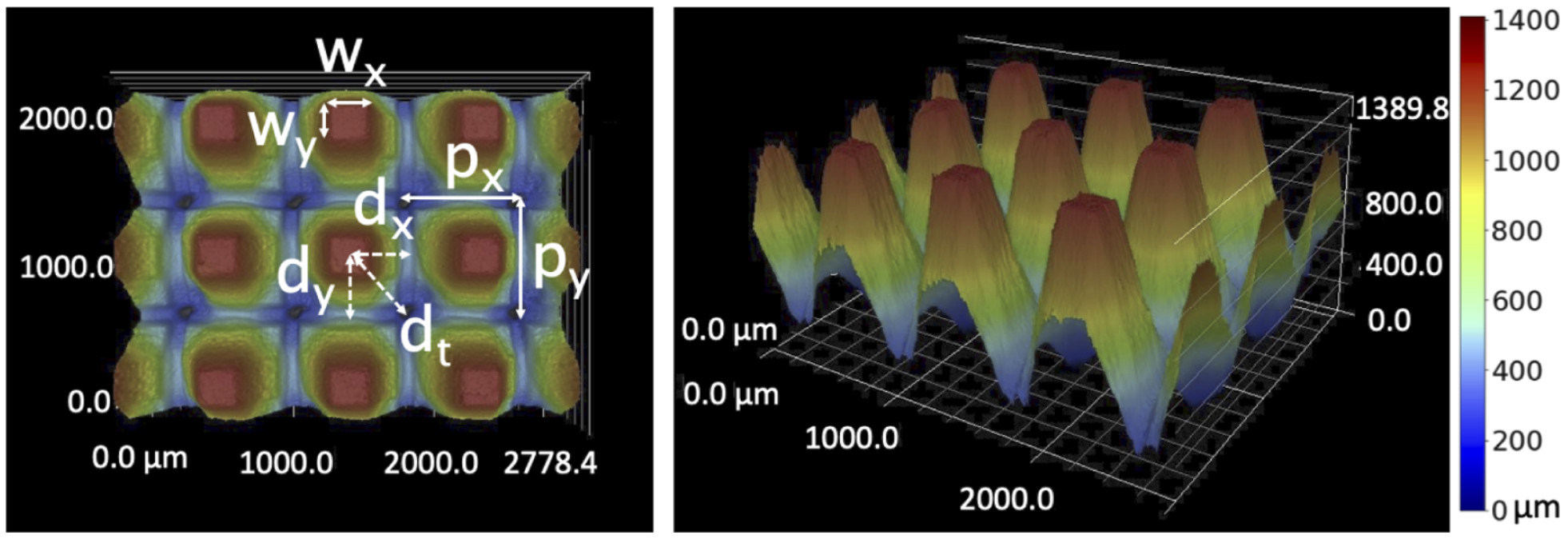

**Fig. 2.** Top and perspective views of a section of the the validation sample. White solid arrows show the definition of shape parameters. Dashed arrows show orientations along which depth measurements are reported in Table 2.

**Table 2. Average pyramid shape parameters for the validation sample and for each of the fabricated filter surfaces in mm. The parameter terminology is annotated in Fig. 2. Error margins indicate standard deviation of measurements on nine pyramids for the validation sample, and on 740 pyramids per side for the fabricated filter.**

| Shape Parameter | Validation sample | Filter side 1 | Filter side 2 |
|---|---|---|---|
| Pitch $x$ ($p_x$) | 0.77 ± 0.01 | 0.77 ± 0.01 | 0.77 ± 0.01 |
| Pitch $y$ ($p_y$) | 0.77 ± 0.01 | 0.77 ± 0.01 | 0.77 ± 0.01 |
| Top width $x$ ($w_x$) | 0.16 ± 0.01 | 0.18 ± 0.01 | 0.19 ± 0.01 |
| Top width $y$ ($w_y$) | 0.18 ± 0.02 | 0.18 ± 0.01 | 0.19 ± 0.01 |
| Saddle depth $x$ ($d_x$) | 0.93 ± 0.01 | 0.95 ± 0.03 | 0.93 ± 0.03 |
| Saddle depth $y$ ($d_y$) | 0.91 ± 0.01 | 0.94 ± 0.01 | 0.93 ± 0.02 |
| Total depth ($d_t$) | 1.30 ± 0.01 | 1.34 ± 0.02 | 1.32 ± 0.02 |

## 4. Fabrication

The laser ablation system consists of an ultrafast laser (Light Conversion Carbide-CB3) operated with the parameters given in Table 3, a galvanometer scanner (Scanlab, intelli$SCAN_{se}$14), and a 3-dimensional (3D) translational stage (Kohzu, ZA16A-32F01, and OptoSigma, XA20F-L2501), which hold the samples. By adjusting the compressor inside the laser, the pulse duration is stretched to 4 ps. The spot diameter on the sample is measured by placing a CMOS camera at the focal point [48].

**Table 3. Laser specifications.**

| Model: Carbide CB3 + CBM03-2H-3H | |
|---|---|
| Average power | 40 W |
| Wavelength | 1030 nm |
| Repetition rate | 100 kHz |
| Pulse duration | 4 ps |
| Pulse energy | 400 $\mu$J |
| Spot diameter (1/e$^2$) | 28 $\mu$m |

The 284 mm diameter area was fabricated by tiling two types of smaller square areas, one with 12×12 pyramids and the other with 6×6 pyramids. The smaller areas were only used at the edge of the optical diameter; see Fig. 3. The scanner was used to ablate one tile-sized area and the 3D stage was used to translate the sample for ablation of different tiles. The laser beam scan pattern at each tile followed repeated passes in $x$ and $y$ similar to the type we used in past fabrications [11,23,24,27]. Because the fabricated structures show a small $x, y$ asymmetry, the sample was rotated by 90 degrees when the second surface was machined.

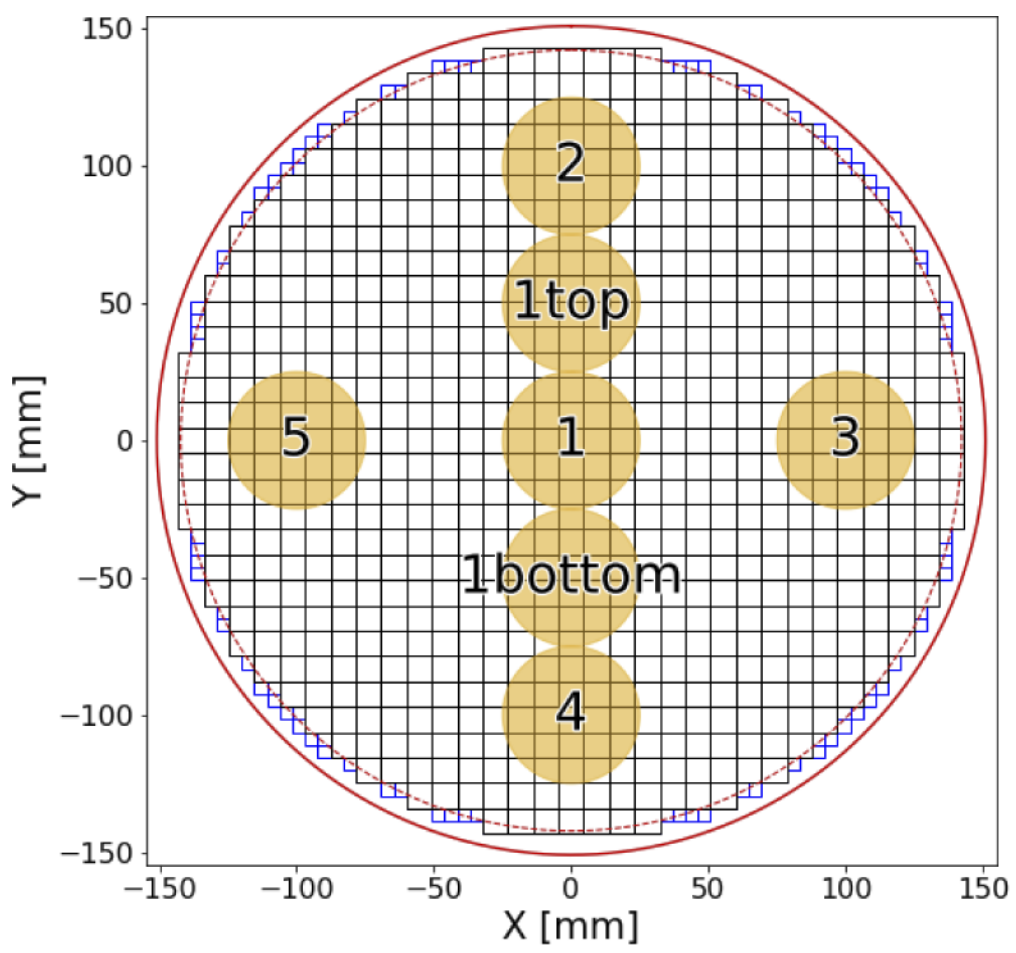

**Fig. 3.** Schematic of the MUSTANG2 filter with key dimensions and definitions of measurement areas (yellow circles). SWS were fabricated across the 284 mm diameter optically active area (red dash circle) by tiling two types of square sub-areas. One with a side of 9.24 mm and 12×12 pyramids (black squares), and the other with a side of 4.62 mm and 6×6 pyramids (blue squares at the edge). The outer diameter of the sample is 302 mm (red solid circle). Confocal microscopy shape measurements were conducted on 90 pyramids in each of areas 2-5, and on 190 pyramids in each of 1top and 1bottom. Millimeter wave transmission measurements of the pre- (post-)ablated sample were conducted through 5 (6) areas: locations 1-5 (locations 2-5 and 1top and 1bottom).

Before fabrication, we aligned the sample for flatness using a Keyence SI-F-10 distance finder, which was mounted on the same optical bench as the laser scanner. By shimming the sample we reduced distance variations along the 300 mm sample diameter to less than 60 $\mu$m, placing an upper limit of 0.01 deg on sample tilt. This maximal tilt was along the line connecting locations 3 and 5 – see Fig. 3 – with location 5 being further away from the scanner when side 1 was ablated.

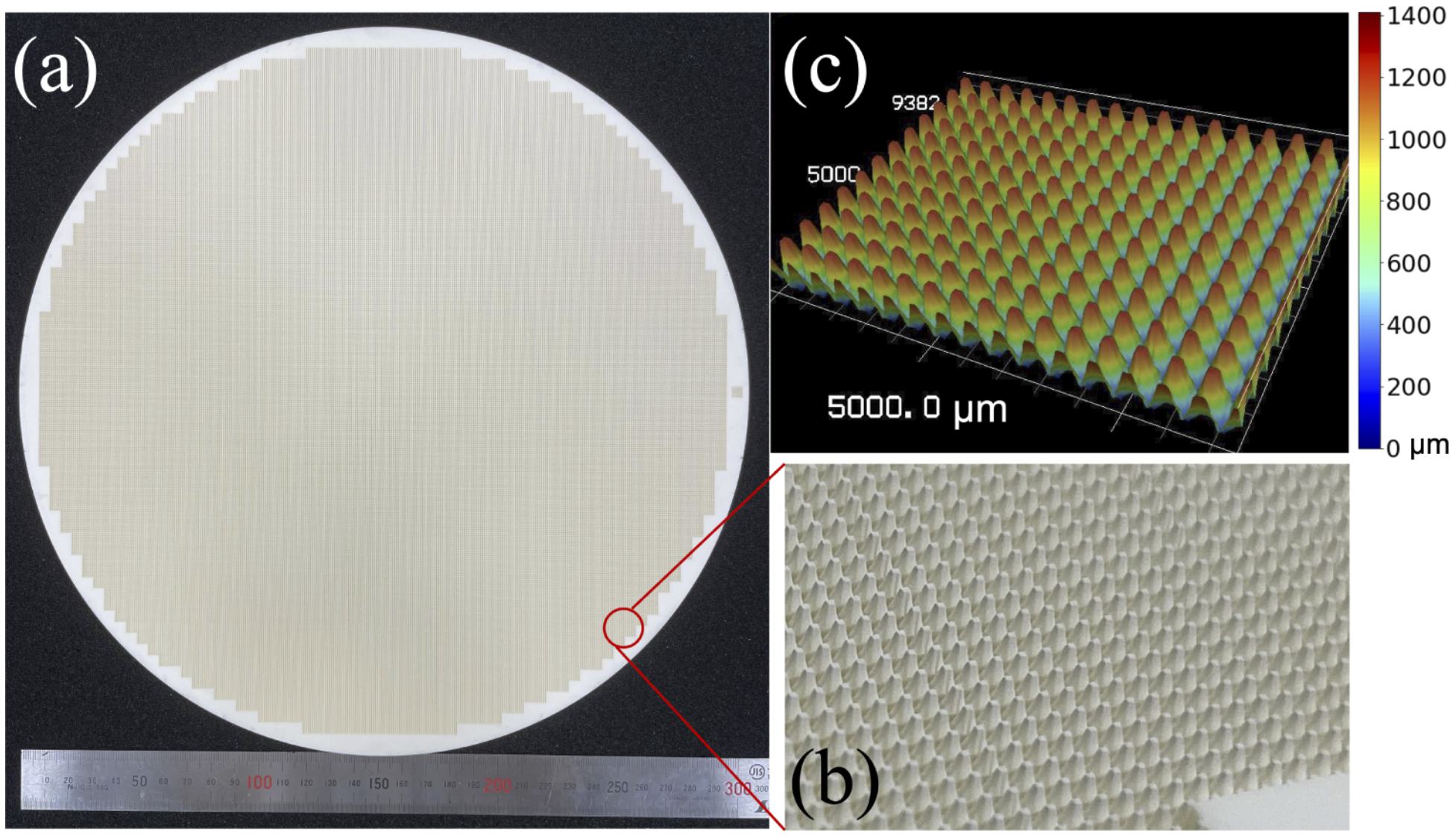

**Fig. 4.** Images of one side of the fabricated filter. The other side is identical. (a) A photograph of the entire filter. The ruler is graduated up to a length of 300 mm. (b) An enlarged area. (c) Rendering of confocal microscopy scanning of the SWS.

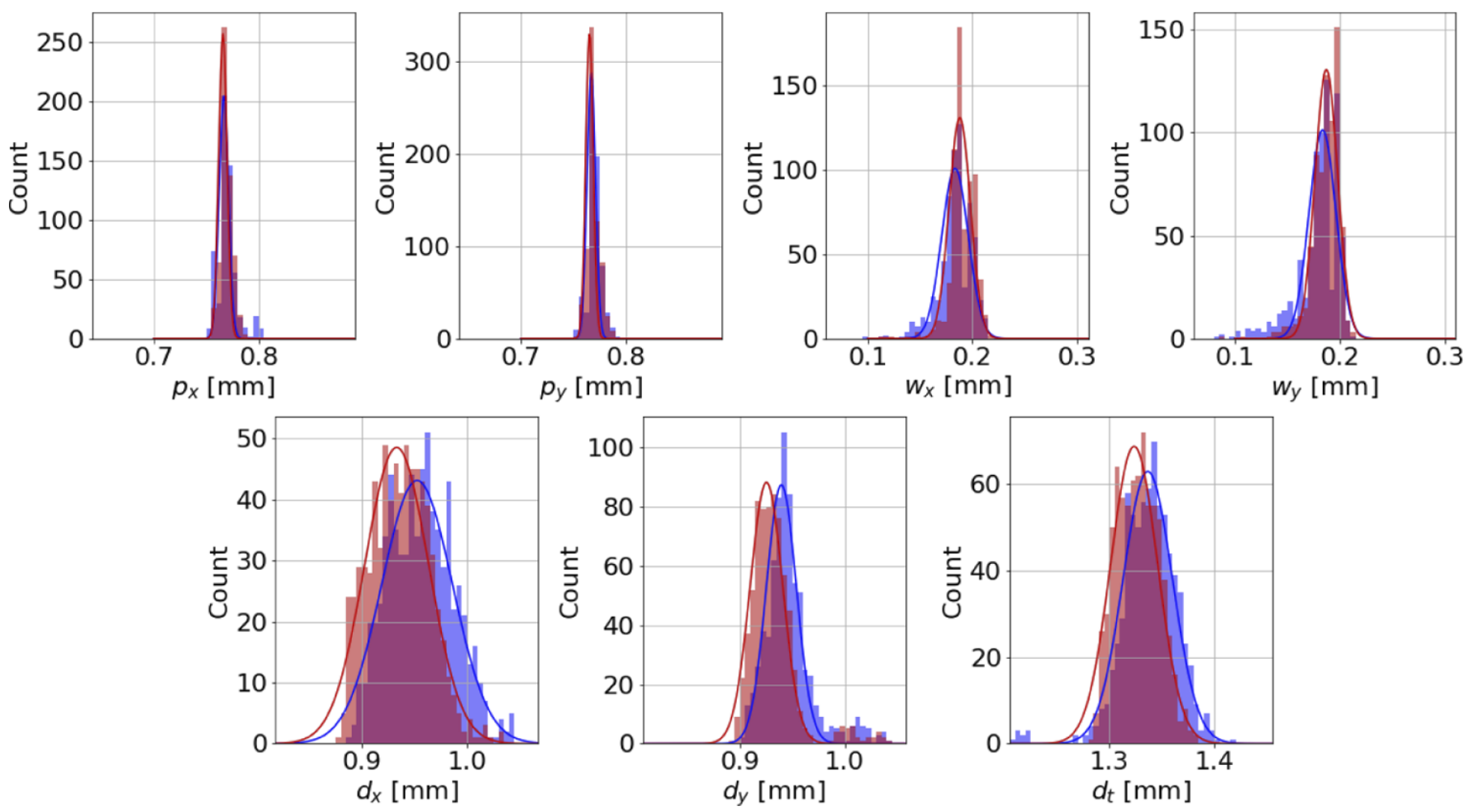

**Fig. 5.** Histograms of shape parameters for 1480 pyramids, 740 for each of side1 (blue) and side2 (red) of the filter. All panels have a horizontal span of 250 $\mu$m to illustrate which parameters have larger variance. The solid curves are best-fit Gaussians. The definition of the shape parameters are given in Fig. 2, and the means and standard deviations of the fits are given in Table 2.

Figure 4 shows an image of the fabricated filter, an enlargement, and a 3D rendering of confocal microscopy scanning. The enlargement and the confocal microscopy cover an area that is larger than a single tile and thus show the quality of the interface between individually machined tiles.

It took 41.5 hours to fabricate each side with 92% of the time spent ablating material and 8% spent non-ablating, primarily steering the beam or moving the sample between ablation passes. The fabrication was automated and done continuously without interruption. Given that 44910 mm$^3$ of material was ablated from each side, the process rate was 18.0 mm$^3$/min and volume ablation rate was 19.1 mm$^3$/min resulting in a process efficiency of 94% [25].

## 5. Measurements

The pre-ablated sample had a diameter of 302 mm and thickness of 5.055$\pm$0.004 mm. We measure the shapes of the fabricated SWS, and conduct millimeter-wave transmission measurements of the pre-ablated sample and of the fully fabricated sample. Millimeter-wave measurements of the pre-ablated sample are used to determine its index of refraction and loss tangent. All measurements are done at room temperature.

### *5.1. Shape parameters*

We use confocal microscopy to image six areas in each side of the machined filter. The locations, 1top, 1bottom, and 2 through 5, are shown in Fig. 3. We measure shape parameters for a total of 1480 pyramids, 740 in each side with 190 pyramids in each of locations 1top and 1bottom, and 90 pyramids in each of locations 2 through 5. Histograms with the results are given in Fig. 5, and means and standard deviations of best-fit Gaussian are given in Table 2. Histograms for each of the measured locations for each side are given in Appendix A.2.

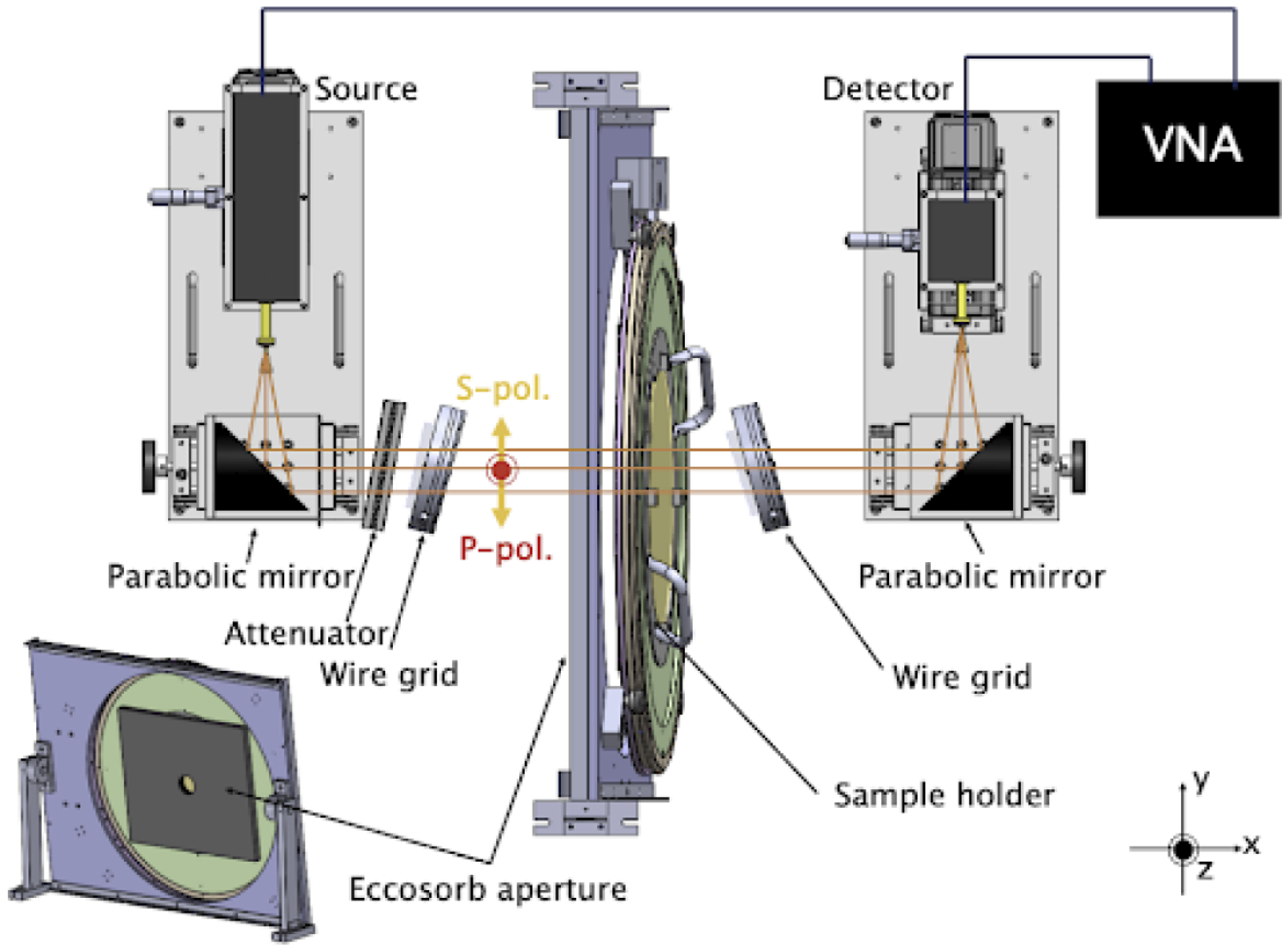

**Fig. 6.** Schematic image of the transmittance measurement setup. To reduce standing waves the attenuator and wire grids are rotated around the z axis by angles between 5-10 degrees, and the sample is rotated by 7 degrees around the y axis. Panel (a) shows the AN73 absorber placed parallel to the sample, on its source side, with an aperture defining the center measurement location (location 1). Similar sheets with holes placed elsewhere defined other measurement locations.

### 5.2. *Millimeter-wave measurement setup and process*

Transmission measurements between 55 and 140 GHz are conducted with a Keysight N5222B vector network analyzer coupled to two parabolic mirrors and two grids, as shown in Fig. 6. We place the sample in the collimated part of the beam. The sample holder can translate relative to the mm-wave beam and the sample can rotate within the holder. To reduce the effects of systematic uncertainties arising from standing waves, we tilt the sample by 7 degrees about the $y$ axis. The grids' transmission axes are aligned to within 0.5 degrees relative to each other. Their co-polarization and cross-polarization transmission is higher than 99% and less than 0.6%, respectively. When the grid transmission axes are horizontal (vertical) the polarization incident on the sample is an s-state (p-state). The wire grids and attenuators made of Eccosorb HR-10 are tilted by 5-10 degrees around the $z$ axis to further reduce standing waves.

Independent measurements are conducted in several 50 mm diameter locations that are labeled in Fig. 3. The measurement location is defined by placing an 18 mm thick sheet of Eccosorb AN-73 absorber on the source side of the sample. The different location of a 50 mm diameter hole in each of several sheets defines the measurement location. Transmission through the absorber itself was measured to be decreasing with increasing frequency, and to be less than 0.1% at 55 GHz and less than 0.01% at frequencies above 75 GHz. Transmission along paths around the sample holder is less than 0.01% at all frequencies.

All transmittance results are obtained by taking the ratio of measured power with the sample to the measured power with an identical setup but without the sample.

### 5.3. *Measurement of index of refraction and loss*

We measured the p-polarization state transmission of the pre-antirefelction coated sample to determine its index of refraction and loss tangent. The transmittance is measured with resolutions of 0.035 and 0.05 GHz in the E-band (55 - 90 GHz), and F-band (90 - 140 GHz), respectively, through locations 1-5, see Fig. 3. The transmission data from all locations is combined, averaged, and $\chi^2$-fit with a TMM-based prediction; see Fig. 7. The index of refraction $n$ and loss tangent $\delta$ are free parameters. The best fit values and uncertainty intervals derived from the $\chi^2$ fit are given in Table 4. We search for frequency dependence of $n$ and $\tan\delta$ by repeating the analysis in the lower and upper frequency bands but do not find any statistically significant change. We repeat the analysis in each of the measured locations and find values for $n$ and $\delta$ that are consistent within uncertainties with the combined data. The individual values of $n$ range from 3.121 to 3.125 and all loss tangents are less than $4.6 \cdot 10^{-4}$.

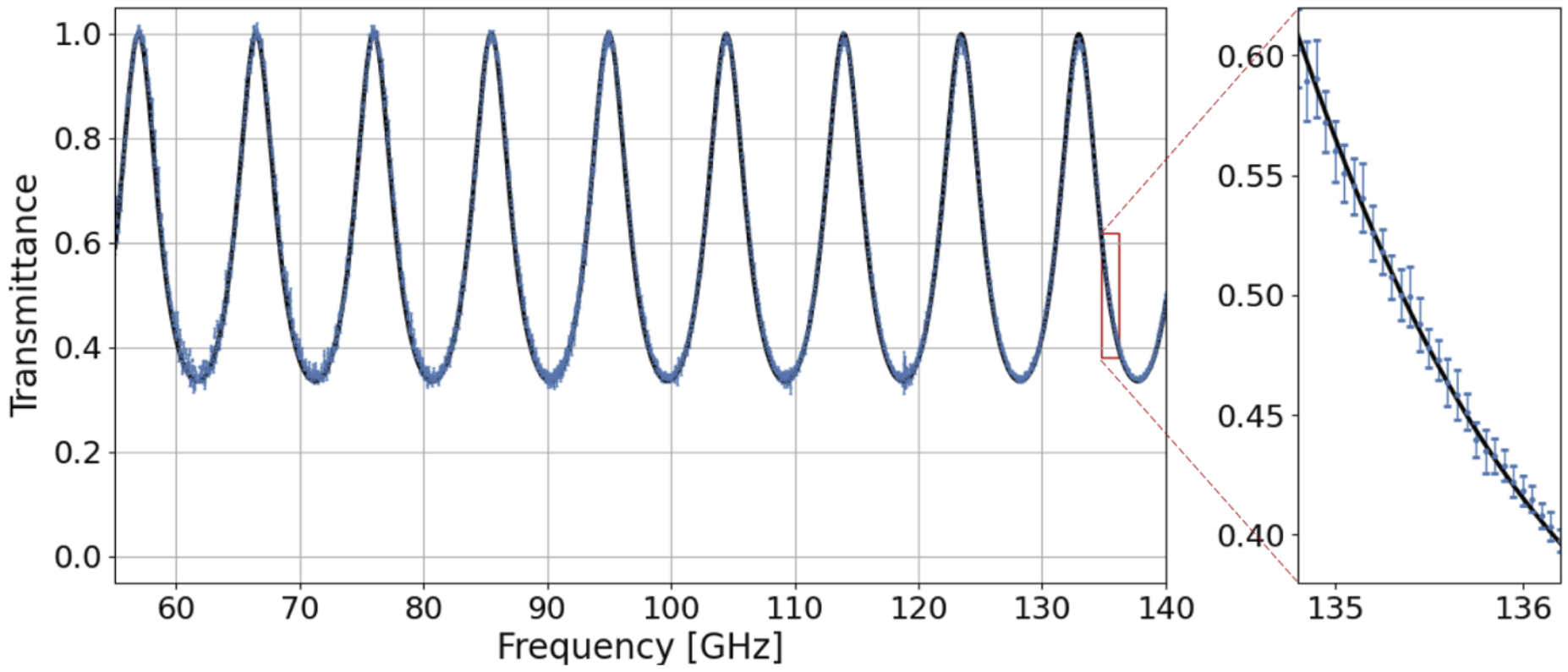

**Fig. 7.** Average transmittance of the filter pre-ARC (left, blue points), the best fit TMM-based model (black line), and an enlargement of a small frequency range between 135 and 136 GHz (right). Error bars are standard deviations of the averaged data.

**Table 4. Measurements of refractive index and loss tangent of the pre-ablated sample.**

| $n$ | $\tan\delta$ |
|---|---|
| 3.123 ± 0.03 (68%) | $<4.6 \times 10^{-4}$ (95%) |

## 5.4. *Filter Transmission*

We measure the transmittance with p- and s-polarization states between 55 and 140 GHz at six locations as shown in Fig. 3 with the same frequency resolution used for the pre-ablated sample transmittance; see Section 5.3. To produce a transmission spectrum per polarization state the data from all locations are combined, binned in 0.5 GHz bins, and averaged in each bin. The uncertainty assigned is the standard deviation of the 84 and 60 data per bin in the E- and F-bands, respectively. To produce a transmission spectrum for unpolarized light the transmission spectra of p- and s-polarization data are averaged and uncertainty propagated. The transmission spectra for the MUSTANG2 frequency band are shown in Fig. 8. We compare the measured data to RCWA transmittance calculations in which we used the average measured pyramid shape for each side of the filter as given in Table 2, an index $n = 3.123$ and $\tan\delta = 4 \cdot 10^{-4}$. We use this value for $\delta$ because (i) it is consistent with measurement with the pre-ablated sample (see Section 5.3) and (ii) it minimizes the $\chi^2$ between the data and the RCWA model for frequencies between 55 and 115 GHz. We do not use higher frequency data to assess the $\chi^2$ because of the onset of diffraction; see Appendix A. The measured and predicted average transmittances between 75 and 105 GHz is 98% for unpolarized light and in each of the polarization states. The difference between the two polarization states is less than 1%.

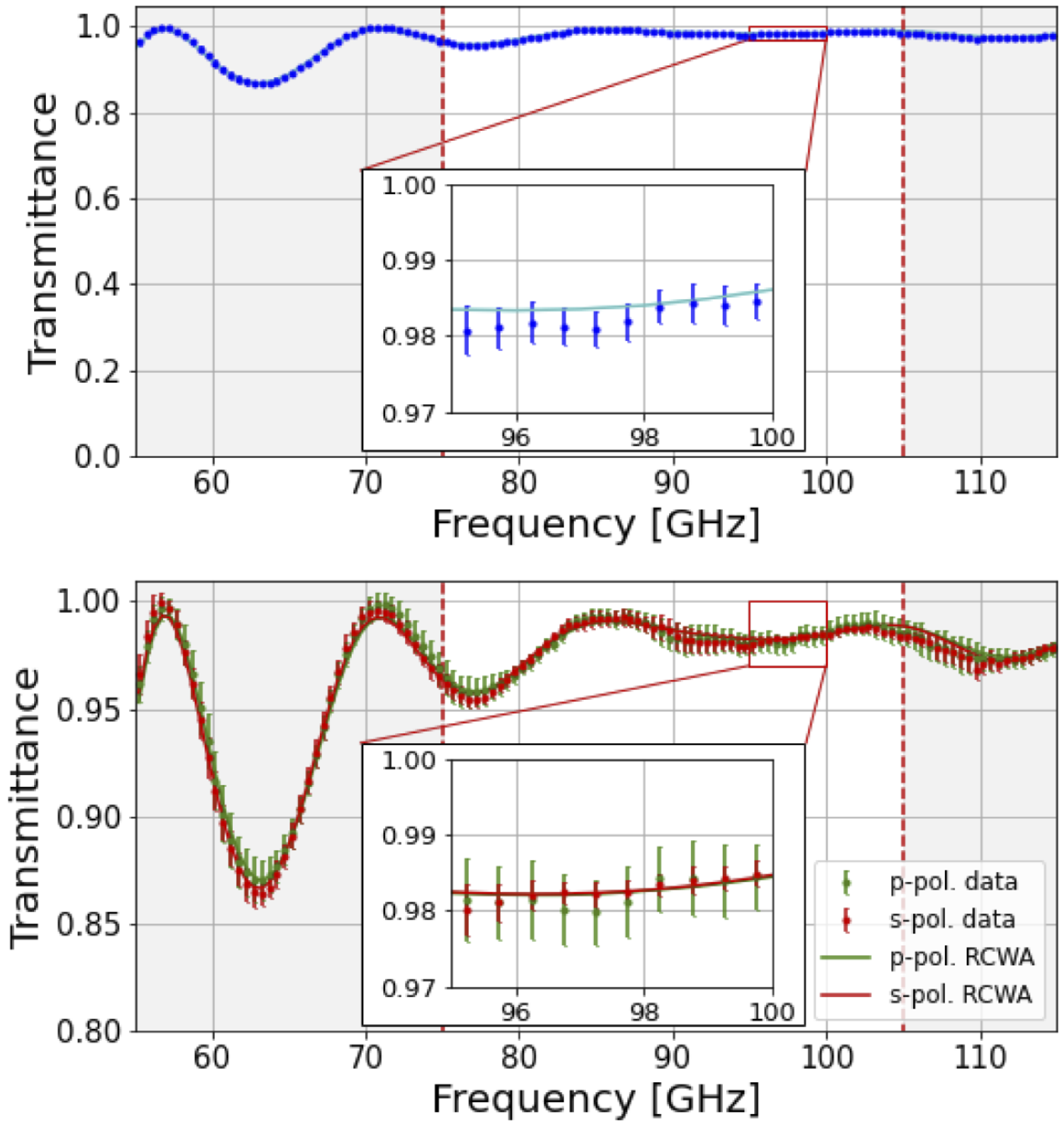

**Fig. 8.** The transmission spectrum for the MUSTANG2 filter for unpolarized light (upper panel, blue dots), for s- and p-polarized light (lower panel, red and green dots, respectively), and RCWA predicted transmission given the average shape data for each side as provided in Table 2 (upper panel, solid cyan, and lower panel solid red and green). The average transmission within the band (white region) is 98%. Note the different vertical spans for the two panels, which provide both an overall performance view and details on the agreement between the data and the RCWA predictions.

The measurements of transmission at each of the six measured locations give results that are consistent with the combined averaged data with very little variance, see Table 5 in the Appendix. Averaged measured and RCWA-calculated transmittance per location range between 97.6% and 98.6% including both polarization states. The RCWA calculations use the average shape measured per location (see Table 6 in the Appendix).

## 6. Discussion

The MUSTANG2 alumina-made low pass filter has been temperature-cycled in the laboratory from 300 to 77 K and back with no adverse effects. It has been integrated into the instrument and is operating at 40 K since August 2021. We find a reduction of 1-2 $\mu$W in optical loading on the 300 mK stage, transforming filters-induced load from a major to minor contributor, and a reduction of ~50% in filters-induced optical load on the detectors. Both changes are consistent with predictions; see Section 2. The latest MUSTANG2 observations of astrophysical sources rule out reductions of optical power due to unusually high absorption along the optical train or due to other loss of efficiency.

The reduction in filters-induced in-band detector optical load translates to a reduction of 10%–40% in the total optical load on the detectors, a range that brackets the worst to best atmospheric condition, respectively; atmospheric optical power is a significant contributor to the total load on the MUSTANG2 detectors under all observation conditions. Had the instrument been operating in Chile or the South Pole, where atmospheric optical load is smaller than in Green Bank, the replacement from Teflon to alumina would have made an even more substantial change in detector optical loading.

The majority of shape parameters have variance of 10 $\mu$m indicating a high level of uniformity among fabricated structures. Such uniformity is a function of laser fluence stability, a property of most modern lasers, and of initial alignment of the sample. The structure uniformity is further supported by the ~1% spatial variance in the measured and predicted band-average transmission. The structure shape variance that does exist, up to 30 $\mu$m for one standard deviation in depth measurements (see Table 2), has an effect of 1% at most on the band-average transmission. Similar conclusions hold for transmittances in the s- and p-polarization states: the difference in band-averaged transmittance between the two states is less than 1%. We did not make any special attempts to further constrain the difference because MUSTANG2 is not polarization sensitive.

The MUSTANG2 focal plane couples to the Green Bank Telescope with an effective focal ratio of 1.94 giving incidence angles of up to 14.5 degrees. The transmittance data was collected at an incidence angle of 7 degrees; see Section 5.2. RCWA calculations using the measured shapes indicate a reduction of less than 0.2% in transmittance for an incidence of 14.5 deg, and essentially no change for normal incidence.

The MUSTANG2 SWS were fabricated on a diameter of 284 mm over 41.5 hours for one side with a volume ablation rate of 19.1 mm$^3$/min. Two sides of similar samples with ~0.5 m diameter could therefore be fabricated within about four weeks of laser time, assuming identical laser parameters. As we have shown in this work, modern instrumentation enables automated machining, making laser time and total work time equal. Wen et al. [25] report alumina ablation rate nearly twice higher, up to 34 mm$^3$/min for structure heights of up to 900 $\mu$m. Their measurements were conducted with laser power up to 100 W. Their data and model show that in almost all cases, higher laser power increases the ablation rate and decreases fabrication time. With the rate demonstrated by Wen et al. two sides of a 0.5 m diameter sample can be machined in 16 days.

## 7. Conclusion

We fabricated a 300 mm diameter low pass filter made of alumina with laser-ablated SWS ARC. The filter has transmittance of 98% between 75 and 105 GHz for all required incidence angles. The SWS ARC reduced reflections to an average of 1%, a factor of 45 reduction relative to a 4 mm-thick non-AR coated surface. The transmittance difference between the s- and p-polarization states is less than 1%. There is no observable variance in these results with location across the filter within an accuracy of 1%. The filter has been integrated into the MUSTANG2 instrument, is operating in cryogenic temperatures, and the improvements anticipated due to its high thermal conductance have been realized. This is the first report of an alumina-based optical element with SWS ARC that has been integrated into an operating instrument, and the first demonstration of a large area fabrication of SWS with laser ablation.

## A. Appendix

### *A.1. Transmittance*

Figure 9 gives the transmission measurements in all locations between 55 and 140 GHz. It includes RCWA predictions based on the SWS average shape measured at each location. The onset of diffraction peaks is apparent for frequencies above 120 GHz. Table 5 gives the s- and p-polarization average transmittance calculated per location from the measured data and from the RCWA calculation. The RCWA calculation uses the average shape measured at the given location, see Table 6.

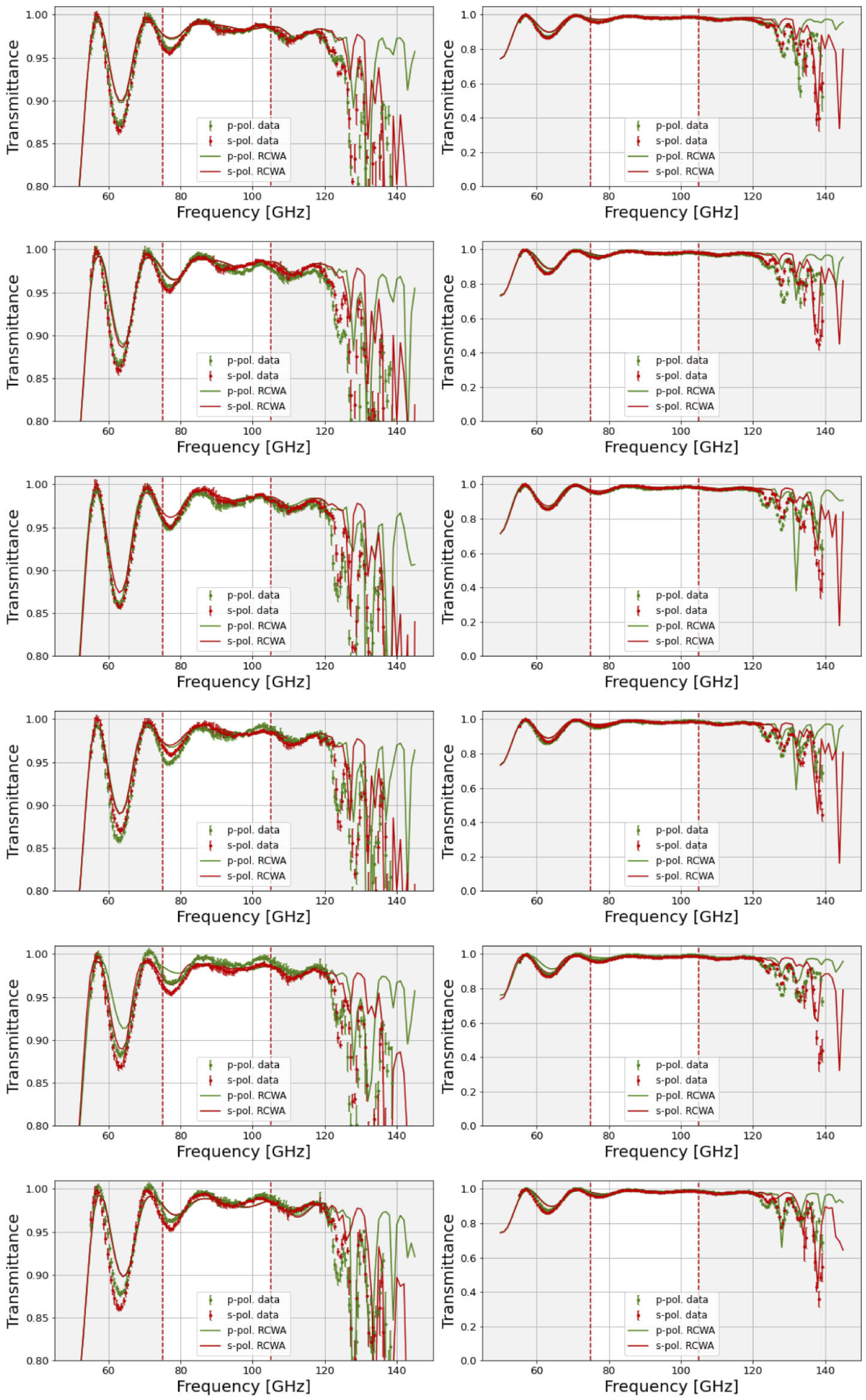

**Fig. 9.** Transmittance measurements between 55 and 140 GHz for p- and s-polarizations (green and red data points, respectively), and RCWA predictions (solid curves) based on the average shape measured at each location. From top to bottom: locations 1top, 1bottom, and 2 - 5. The left and right columns span different transmission values. The MUSTANG2 band is between 75 and 105 GHz (white region). Onset of diffraction is visible above 120 GHz.

**Table 5. Average measured and calculated transmittance per location in s- and p-polarization states.**

| Location | $T_p$ (Measurement) [%] | $T_p$ (Calculation) [%] | $T_s$ (Measurement) [%] | $T_s$ (Calculation) [%] |
|---|---|---|---|---|
| 1top | 97.8 | 97.9 | 97.9 | 97.9 |
| 1bottom | 98.1 | 98.2 | 98.1 | 98.2 |
| 2 | 97.6 | 97.7 | 98.0 | 97.9 |
| 3 | 97.9 | 98.1 | 98.2 | 98.2 |
| 4 | 98.6 | 98.2 | 97.9 | 97.9 |
| 5 | 98.4 | 98.0 | 98.1 | 98.0 |
| Average | 98.1 | 98.0 | 98.0 | 98.0 |

**Table 6. Averaged pyramid shape parameters for the fabricated filter at each location. The uncertainty reflects the standard deviation of the *N* pyramids measured. Parameter definitions are identical to Table 2.**

| Location ($N$) | Parameters | Side 1 (mm) | Side 2 (mm) |
|---|---|---|---|
| 1top (190) | $p_x$ | 0.77 ± 0.01 | 0.77 ± 0.01 |
| | $p_y$ | 0.77 ± 0.01 | 0.76 ± 0.01 |
| | $w_x$ | 0.18 ± 0.01 | 0.18 ± 0.01 |
| | $w_y$ | 0.19 ± 0.01 | 0.19 ± 0.01 |
| | $d_x$ | 0.95 ± 0.03 | 0.92 ± 0.03 |
| | $d_y$ | 0.94 ± 0.01 | 0.92 ± 0.01 |
| | $d_t$ | 1.34 ± 0.02 | 1.31 ± 0.02 |
| 1bottom (190) | $p_x$ | 0.77 ± 0.01 | 0.77 ± 0.01 |
| | $p_y$ | 0.77 ± 0.01 | 0.77 ± 0.01 |
| | $w_x$ | 0.18 ± 0.01 | 0.18 ± 0.01 |
| | $w_y$ | 0.18 ± 0.01 | 0.19 ± 0.01 |
| | $d_x$ | 0.97 ± 0.03 | 0.96 ± 0.03 |
| | $d_y$ | 0.95 ± 0.01 | 0.94 ± 0.01 |
| | $d_t$ | 1.36 ± 0.02 | 1.34 ± 0.01 |
| 2 (90) | $p_x$ | 0.77 ± 0.01 | 0.77 ± 0.01 |
| | $p_y$ | 0.77 ± 0.01 | 0.77 ± 0.01 |
| | $w_x$ | 0.19 ± 0.01 | 0.18 ± 0.01 |
| | $w_y$ | 0.19 ± 0.01 | 0.19 ± 0.01 |
| | $d_x$ | 0.91 ± 0.01 | 0.94 ± 0.02 |
| | $d_y$ | 0.89 ± 0.01 | 0.90 ± 0.00 |
| | $d_t$ | 1.31 ± 0.02 | 1.33 ± 0.02 |
| 3 (90) | $p_x$ | 0.77 ± 0.01 | 0.77 ± 0.01 |
| | $p_y$ | 0.77 ± 0.01 | 0.77 ± 0.01 |
| | $w_x$ | 0.20 ± 0.01 | 0.20 ± 0.01 |
| | $w_y$ | 0.18 ± 0.01 | 0.17 ± 0.01 |
| | $d_x$ | 0.93 ± 0.01 | 0.94 ± 0.02 |
| | $d_y$ | 0.94 ± 0.01 | 0.94 ± 0.01 |
| | $d_t$ | 1.34 ± 0.02 | 1.35 ± 0.01 |
| 4 (90) | $p_x$ | 0.77 ± 0.01 | 0.77 ± 0.01 |
| | $p_y$ | 0.77 ± 0.01 | 0.77 ± 0.01 |
| | $w_x$ | 0.18 ± 0.01 | 0.18 ± 0.01 |
| | $w_y$ | 0.18 ± 0.01 | 0.19 ± 0.01 |
| | $d_x$ | 0.98 ± 0.03 | 0.92 ± 0.03 |
| | $d_y$ | 0.95 ± 0.01 | 0.89 ± 0.01 |
| | $d_t$ | 1.37 ± 0.02 | 1.32 ± 0.02 |
| 5 (90) | $p_x$ | 0.77 ± 0.01 | 0.77 ± 0.01 |
| | $p_y$ | 0.77 ± 0.01 | 0.77 ± 0.01 |
| | $w_x$ | 0.20 ± 0.01 | 0.20 ± 0.01 |
| | $w_y$ | 0.18 ± 0.01 | 0.18 ± 0.01 |
| | $d_x$ | 0.96 ± 0.01 | 0.92 ± 0.02 |
| | $d_y$ | 0.94 ± 0.01 | 0.91 ± 0.01 |
| | $d_t$ | 1.36 ± 0.01 | 1.30 ± 0.01 |

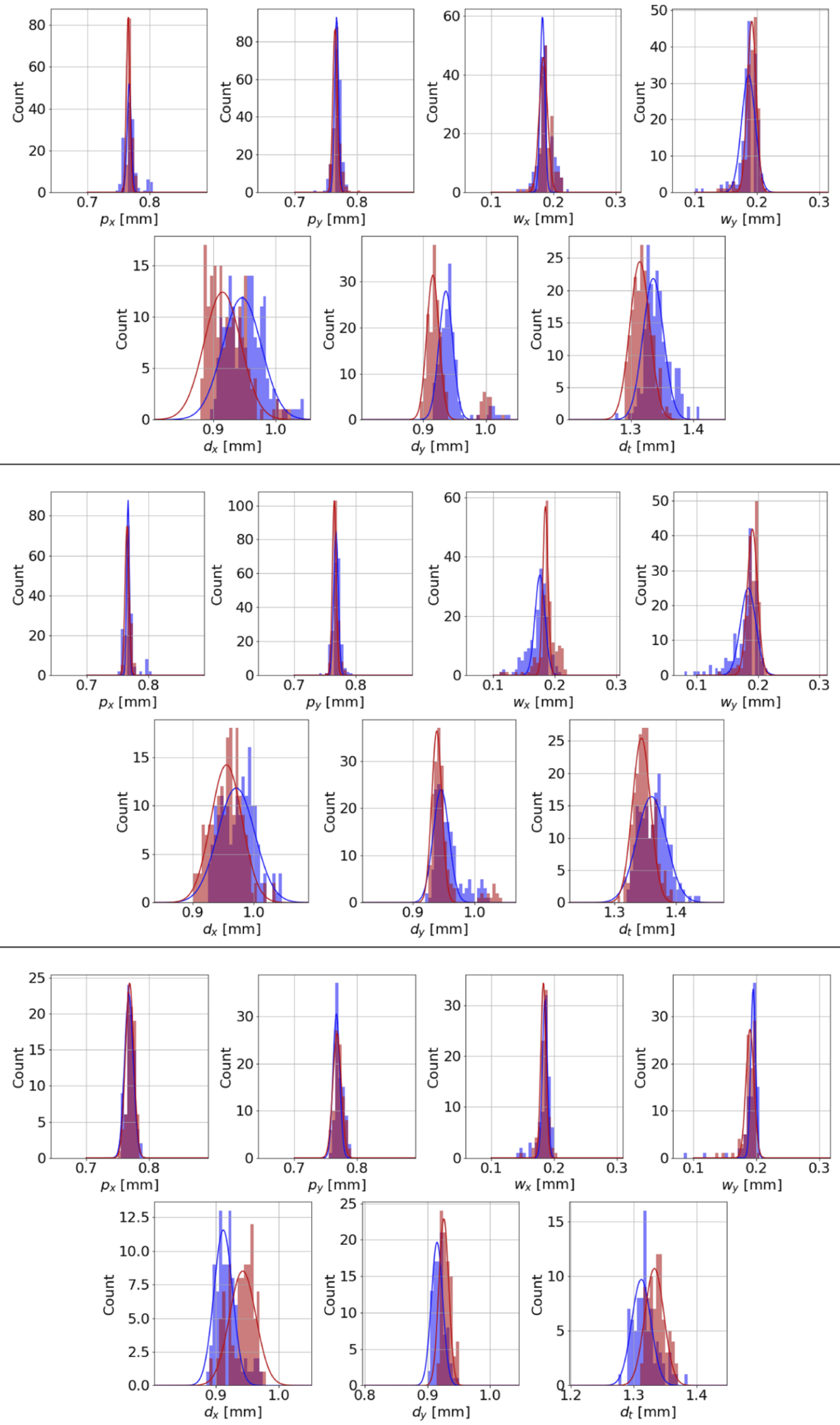

**Fig. 10.** Histograms of shape parameter measurements for locations 1top (top), 1bottom (middle) and location 2 (bottom), for side1 (blue) and side2 (red), and best fit Gaussians (solid lines).

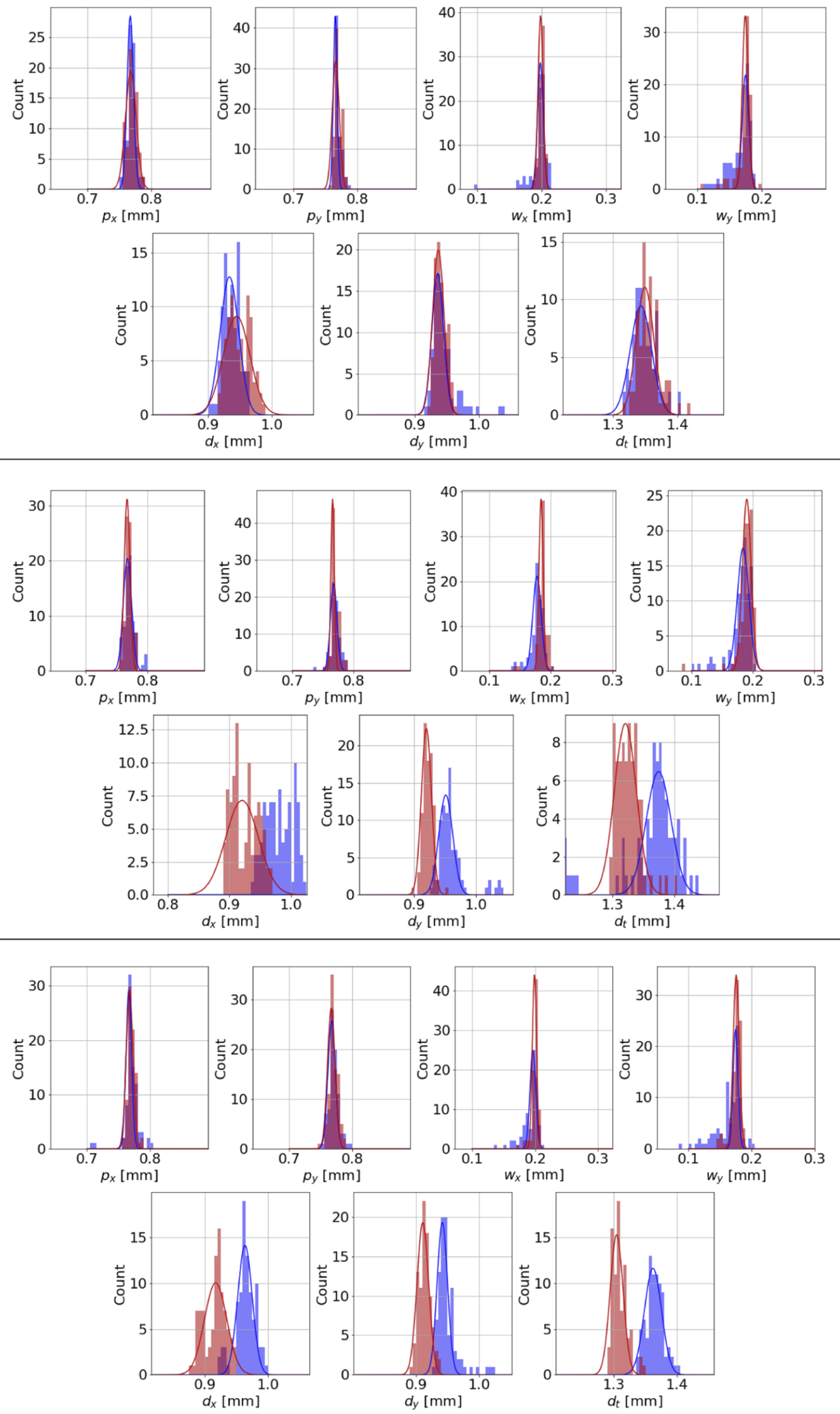

**Fig. 11.** Same as in Fig. 10 for location 3 (top), location 4 (middle), and location 5 (bottom).

### A.2. Shape measurements at six independent locations

Table 6 gives the average SWS shape parameters for each of the six locations as described in Fig. 3. The data in the table was combined to produce the averages and standard deviations given in Table 2. Figures 10 and 11 give histograms showing the distributions of the measured parameter for each location.

**Funding.** Japan Science and Technology Agency (JPMJCE1313, JPMXS0118067246); National Science Foundation (ECCS-2025124); Japan Society for the Promotion of Science (18KK0083); Council for Science, Technology and Innovation (P16011).

**Acknowledgments.** We acknowledge the World Premier International Research Center Initiative (WPI), MEXT, Japan, for support through Kavli IPMU.

This work was supported by the New Energy and Industrial Technology Development Organization (NEDO) project "Development of advanced laser processing with intelligence based on high brightness and high efficiency laser technologies" (P16011) by the Council for Science, Technology and Innovation (CSTI), Cross-ministerial Strategic Innovation Promotion Program (SIP), "Photonics and Quantum Technology for Society 5.0", and by the Center of Innovation Program, from Japan Science and Technology Agency, JST (JPMJCE1313). This research was also supported by MEXT Q-LEAP (Grant No. JPMXS0118067246) and JSPS KAKENHI Grant Numbers 18KK0083.

Portions of this work were conducted in the Minnesota Nano Center, which is supported by the National Science Foundation through the National Nanotechnology Coordinated Infrastructure (NNCI) under Award Number ECCS-2025124.

**Disclosures.** The authors declare no conflicts of interest.

**Data availability.** Data underlying the results presented in this paper will be made available in the Data Repository for the University of Minnesota [49].